\documentclass[12pt,epsfig]{article}
\usepackage{graphicx}
\usepackage{amsfonts}
\usepackage[mathscr]{eucal}
\usepackage{latexsym}
\usepackage{epsfig}

\def\be{\begin{equation}}
\def\ee{\end{equation}}
\def\ba{\begin{eqnarray}}
\def\ea{\end{eqnarray}}

\begin{document}
\baselineskip=15.5pt
\pagestyle{plain}

\rightline{NUB-3213/Th-01}
\rightline{hep-ph/0104114}

\begin{center}

\vskip 1.7 cm

{\Large \bf Hadronic interactions, precocious unification, and cosmic ray 
showers at Auger energies}\\


\vskip 2.5cm
Luis Anchordoqui, Haim Goldberg, Jared MacLeod, Tom McCauley, 
Tom Paul,\\ Steve Reucroft, and John Swain

\medskip
\medskip
{\it Department of Physics, Northeastern University\\
Boston, MA 02115, USA}

\vspace{2cm}

\end{center}

\noindent

At Auger energies only model predictions enable us to extract 
primary cosmic ray features. The simulation of 
the shower evolution depends sensitively on the first few interactions,
necessarily related to the quality of our understanding of high energy 
hadronic collisions. Distortions of the standard ``soft semi-hard'' scenario
include novel large compact dimensions and a string or 
quantum gravity scale not far above the electroweak scale. 
Na\"{\i}vely, the additional degrees of freedom yield unification of 
all forces in the TeV range. In this article we
study the influence of such precocious unification during 
atmospheric cascade developments by analyzing the most relevant 
observables in proton induced showers.

\vspace{3.5cm}




\newpage

\tableofcontents

\section{Introduction}

\hfill

Very recently, it has become evident that 
a promising route towards reconciling the apparent mismatch of 
the fundamental scales of particle physics and gravity is to modify 
the short distance behavior of gravity at scales much larger than 
the Planck length. Such modification can be most simply achieved 
by introducing extra dimensions (generally thought to be 
curled-up) in the sub-millimiter range \cite{ADD,rs}. Within this framework 
the fundamental scale of gravity $M_*$ can be lowered all the way 
to ${\cal O}$ (TeV), and the 
observed Planck scale turns out to be just an effective scale valid for 
energies below 
the mass of Kaluza--Klein (KK) excitations \cite{kk}. Clearly, while the 
gravitational 
force has not been directly measured far below the millimeter 
range \cite{hoyle}, Standard 
Model (SM) interactions have been investigated well below this scale. 
Therefore, if large extra dimensions really exist, one needs some mechanism 
to prevent 
SM particles from feeling those extra dimensions. Remarkably, there 
are several possibilities to confine SM fields (and even gravity) to a 4 
dimensional subspace (referred to as a brane-world) within the $(4+n)$  
dimensional spacetime \cite{trapping}. While the phenomenology of $n$ large 
compact dimensions and TeV scale strings is very exciting on its 
own \cite{pheno}, 
at the same time it opens up new scenarii in which to explore ``exotic'' 
KK-cosmologies \cite{cosmology}, as well as extraordinary astrophysical 
effects \cite{astro}. 
Naturally, an intense activity to assess its experimental validity in 
collider experiments is currently underway \cite{exp}.

The extremely high center-of-mass (c.m.) energies attained in 
cosmic ray collisions at the top of the atmosphere are well above 
those necessary to excite the hypothetical KK modes which would reflect a
change in spacetime dimensionality. Therefore, a natural question to ask is 
wether KK excitations could have a direct influence in the development of 
extensive air showers. Cascades initiated by neutrinos (with cross 
sections reaching typical hadronic values) have been extensively discussed 
elsewhere \cite{nussinov,nu,guenter,michael,jain}.  
We concentrate here on proton-induced showers.   
Before going into the technical details of the shower simulations, 
let us summarize the pecularities of gravity with $n>0$ 
that can ruffle the standard ``soft semi-hard'' scenario. 

\hfill

\section{Back of the Envelope Insights into KK-Modes Phenomenology}

\hfill

In the canonical example of \cite{ADD}, spacetime is a direct product of 
ordinary four-dimensional spacetime and a (flat) spatial $n$-torus
with circumferences $L_i = 2 \pi r_i$ ($i=1,\dots,n$), generally of 
common linear size $r_i=r_c$.
As mentioned above, SM fields cannot propagate freely in the extra dimensions
without conflict with observations. This is avoided by trapping the fields 
to a thin shell of thickness $\delta \sim M_s^{-1}$ \cite{g}. The only 
particles propagating in the (4+n) dimensional bulk are the (4+n) gravitons.
Because of the compactification, the extra $n$ components of the 
graviton momenta are quantized
 \begin{equation}
k_i = \frac{2\pi\ell_i}{L_c} = \frac{\ell_i}{r_c}, \,\,\,\, i=1,\dots,n.  
\end{equation}
Thus, taking into account the degeneracy on $\ell_i \in \mathbb{Z}$, 
the graviton looks like a massive KK state with mass 
\begin{equation}
m_{\ell_1, \dots, \ell_n} = \left(\sum_{i=1}^n 
\ell_i^2\right)^{1/2} \, r_c^{-1}.
\label{3}
\end{equation}
It is important to stress that the graviton's self-interactions 
must conserve both ordinary 4-momenta and KK momentum components, whereas 
SM fields (that break 
translational invariance) do not have well 
defined KK momenta in the bulk  for $\ell/r_c \leq M_s$ \cite{ouo}. Therefore,
interactions of gravitons with SM particles do not conserve KK 
momentum components.

Applying Gauss' 
law at $r \ll r_c$ and $r \gg r_c$, it is easily seen that the Planck scale 
of the four dimensional world is 
related to that of a higher dimensional space-time simply by a volume factor,
\begin{equation}
r_c = \left( \frac{M_{\rm pl}}{M_*} \right)^{2/n} \, \frac{1}{M_*} = 2.0 \times 10^{-17} \left(\frac{{\rm TeV}}{M_*}\right)  \left( \frac{M_{\rm pl}}{M_*} \right)^{2/n} \,\,{\rm cm},
\label{1}
\end{equation}
so that $M_*$ can range from $\sim$ TeV to $M_{\rm pl} = 10^{18}$ GeV, 
for $r_c\leq 1$ mm and $n \geq 2$.
For $n\leq 6$, the mass splitting, 
\begin{equation}
\Delta m \sim \frac{1}{r_c} = M_* \left(\frac{M_*}{M_{\rm pl}} \right)^{2/n} \sim
\left(\frac{M_*}{{\rm TeV}} \right)^{n+2/2} 10^{(12\,n-31)/n} \,\,{\rm eV},
\end{equation}
is so small that the sum over the tower of KK states can be replaced by a 
continuous integration. Then the number of modes between $|\ell|$ and 
$|\ell| + d\ell$ reads,
\begin{equation}
dN = d\ell_1 d\ell_2 \dots d\ell_n = S_{n-1} \,|\ell|^{n-1} d\ell, 
\label{states}
\end{equation}
where 
\begin{equation}
S_{n-1} = \frac{2\,\pi^{n/2}}{\Gamma(n/2)}
\end{equation}
is the surface of a unit-radius sphere in $n$ dimensions.
Now using Eqs. (\ref{3}) and (\ref{1}), Eq. (\ref{states}) can be re-written 
as
\begin{equation}
dN  = S_{n-1} \left(\frac{M_{\rm pl}}{M_*}\right)^2 \, \frac{1}{M_*^n} \, m^{n-1}\, dm.
\end{equation}
From the 4-dimensional viewpoint the graviton 
interaction vertex is suppressed by $M_{\rm pl}$. Roughly speaking, 
$\sigma_m \propto  M_{\rm pl}^{-2}$. 
Now, introducing $d\sigma_m/dt$, the differential cross section 
for producing a single mode of mass $m$, one can write down 
the differential cross section for inclusive graviton production 
\begin{equation}
\frac{d^2\sigma}{dt\,dm} = S_{n-1} \left(\frac{M_{\rm pl}}{M_*}\right)^2 \, 
\frac{1}{M_*^n} \, m^{n-1}\, \frac{d\sigma_m}{dt},
\label{zero}
\end{equation}
or else, the branching ratio for emitting any one of  
the available gravitons  
\begin{equation}
\Gamma_g  \sim \frac{s^{n/2}}{M_*^{2+n}},
\label{silence}
\end{equation}
where $s^{1/2}$ is the c.m. energy available for graviton-KK emission.
All in all, one can see by inspection of Eq. (\ref{silence}) that the enormous 
number of accessible KK-states can compensate for the $M_{\rm pl}^2$ factor 
in the scattering 
amplitude.

Having outlined the general ideas for the production of KK 
gravitons in very high energy collisions, let us consider now the 
hadronic scattering of two SM particles. To illustrate the 
effect of extra dimension gravity, we will estimate the effects of 
exchanging a tower of KK gravitons between the hadrons, rather than the 
production of soft gravitons.
As usual, the parton evolution of interacting hadrons $a$ and $b$
must be separated into: (i) the non-perturbative soft cascades, 
characterized by a small momentum transfer $q_t<q_0 \approx 2$ GeV
and described by soft Pomeron exchange, (ii) the hard cascades, $q_t>q_0$,  
that should be described perturbatively \cite{qgsjet}.
If one envisions a scattering process considering the exchange of gravitons,
as a qualitative assessment the overall shape of the cross section 
could be written as \cite{csec}
\begin{equation}
\sigma_{\rm tot} = \sigma^{\rm KK} + \sigma^{4{\rm -dim}} 
\label{cst}
\end{equation}
where $\sigma^{\rm KK}$ denotes the contribution from the virtual 
graviton exchange, and
\begin{equation}
\sigma_{ab}^{4{\rm -dim}}(s) = \frac{1}{C_{ab}} \int d^2b \, \left\{1 - 
e^{-C_{ab}\, [\chi_{ab}^{\rm soft}(s,b)\, + \,\chi_{ab}^{\rm hard}(s,b)]} \right\}.
\end{equation} 
Here, $\chi_{ab}^{\rm soft}(s,b)$ stands for the soft eikonal 
defined by \cite{ter}
\begin{equation}
\chi_{ab}^{\rm soft}(s,b) = \frac{\gamma_a\,\gamma_b}{R_{ab}^2} \,\exp 
\left(\Delta y - \frac{b^2}{4\,R_{ab}^2}\right),
\end{equation}
where $b$ is the impact parameter, $y = \ln s$, $\Delta = \alpha_P(0) - 1$, 
and
$R_{ab}^2 = R_a^2 + R_b^2 + \alpha'_P (0) y$. The parameters of the 
Pomeron trajectory ($\Delta$ and $\alpha'_P(0)$) as well as those 
describing the Pomeron-hadron vertices ($\gamma$ and $R^2$) are set to  
their values in {\sc qgsjet}  in the air shower simulation \cite{qgsjet}. 
The semi-hard interaction is treated as the soft Pomeron emission (soft 
pre-evolution) followed by the hard interaction of partons  
\begin{equation}
\chi_{ab}^{\rm hard}(s,b) = \frac{1}{2}\, r^2 \int dy_1 \int dy_2\,\,
\chi_{ab}^{\rm soft}(e^{y_a+y_b},b)\,\, 
\sigma_{\rm hard}(e^{y-y_a-y_b},q_0),
\end{equation}
where $y_{1(2)}$ are the rapidities of the Pomeron end, $\sigma_{\rm hard}$ is the parton interaction cross section, $r^2$ is an adjustable parameter associated with parton density and $C_{ab}$ is the shower enhancement 
coefficient \cite{kaida}.
The latter is also fixed to the value of {\sc qgsjet} in 
the simulations.

A complete theory of massive KK graviton modes is not yet
available, making it impossible to know the exact cross section at
asymptotic energies. A simple Born approximation to the elastic
cross section leads, without modification, to
$\sigma^{{\rm KK}}\sim s^2$ \cite{jain}. Unmodified, this behavior 
by itself eventually
violates unitarity. This may be seen either by examining the
partial waves of this amplitude, or by noting the high energy
Regge behavior of an amplitude  with exchange of the graviton
spin-2 Regge pole: with  intercept $\alpha(0)=2$, the elastic
cross section
\begin{equation}
\frac{d\sigma}{dt}\, \sim\, \frac{|A_R(s,t)|^2}{s^2}\, \sim 
s^{2\alpha(0)-2}\,\sim s^2,
\end{equation} 
whereas the total cross section
\begin{equation}
\sigma^{{\rm KK}}\, \sim \frac{{\rm Im}[A_R(0)]}{s}\,\sim s^{\alpha(0)-1}\,\sim s,
\end{equation}
so that eventually $\sigma^{\rm KK}_{\rm{el}}>\sigma^{\rm KK}.$ 
Eikonal
unitarization schemes modify these behaviors: in the case of the
tree amplitudes \cite{nussinov} the resulting (unitarized) cross
section $\sigma^{{\rm KK}}\sim s,$ whereas for the single Regge pole exchange
amplitude, $\sigma^{{\rm KK}}\sim \ln^2(s/s_0)$ \cite{michael}. However, the
Regge picture of graviton exchange  is not yet entirely
established: both the (apparently) increasing  dominance assumed
by successive Regge cuts due to multiple Regge pole exchange
\cite{nussinov,muzinich}, as well as the 
presence of
the zero mass graviton can introduce considerable uncertainty in
the eventual energy behavior of the cross section. Hereafter, we
work within the unitarization framework and adopt as our cross 
section \cite{guenter}
\begin{equation}
\sigma^{{\rm KK}} \approx \frac{4 \pi s}{M_*^4} \approx 10^{-28} 
\left(\frac{M_*}{{\rm TeV}}\right)^{-4} \,\, \left(\frac{E}{10^{19}\,
{\rm eV}}\right)\,{\rm cm}^2.
\label{cs}
\end{equation}

\hfill

\section{Air Shower Simulations}

\hfill

The experimental information 
obtained at ground level is only indirectly connected 
to the first few generations of hadrons. 
Consequently, the study of the influence 
of KK-modes on hadronic interactions with c.m. energies $s^{1/2}>100$ TeV, 
requires correctly simulating the intrinsic fluctuations in the air showers.

Let us first discuss in a very general way the  
possible effects introduced by  virtual graviton exchange. 
The survival probability 
$N$ at atmospheric depth $X$ of a particle 
$a$ with mean free path
\begin{equation}
\lambda_a = \frac{m_{\rm air}}{\sigma_{a-{\rm air}}},
\label{po}
\end{equation}
is given by
\begin{equation}
N(X) = e^{-X/\lambda_a},
\label{po2}
\end{equation}
where $m_{\rm air}$ is the mass of an average atom of air \cite{M_air}, 
and the cross sections $\sigma_{a-{\rm air}}$ inferred from Eq. (\ref{cs}) 
are  shown in Fig. 1. It is straightforward to see that the total 
thickness of the atmosphere corresponds to more than 
20 hadronic interaction lengths, depending on the primary zenith angle. 
The key feature in the evolution of the shower is the 
branching between decay and interaction of secondary hadrons
along their path in the atmosphere. The latter strongly depends 
both on particle energy and target density. 

Because of the low air density at the top of the atmosphere 
the point of the first interaction fluctuates considerably from 
shower to shower. 
However, KK-graviton exchange significantly reduces the nucleon 
attenuation length, e.g., at $3 \times 10^{20}$ 
eV, $\lambda^{4}_p \approx 41$ g/cm$^2$, whereas $\lambda^{(4+n)}_p 
\approx 38$ g/cm$^2$.
Moreover, tiny deviations on the mean free path of non-leading secondaries 
yield a small change in the shower interaction length. Namely, 
the survival probability of a secondary 
pion (say $E = 5 \times 10^{19}$ eV) at $X = 40$ g/cm$^2$ is reduced  
from 43\% to 41\%, and that of a kaon with the same energy from 30\% to 29\%. 
Therefore, one can -- perhaps na\"{\i}vely -- state that 
phenomenological models 
considering the virtual exchange of graviton towers would trigger, on average, 
earlier shower developments than a naked ``soft semi-hard'' scenario. 

Test simulations runs of giant air shower evolution have been performed, 
choosing typical parameters for the experimental situation at the Fly's Eye 
and Auger experiments \cite{auger}. The algorithms of 
{\sc aires} (version 2.1.1) \cite{sergio} were slightly modified so as to 
track the particles in the atmosphere via the standard 8 parameter 
function,
\begin{equation}
\lambda_a = P_1 \,\frac{ 1 + P_2\,\, u + P_3 \,\,u^2 + P_4\,\,u^3}{1 
+ P_5 \,\,u 
+ P_6 \,\,u^2 + P_7 \,\, u^3 + P_8 \,\,u^4} \,\, {\rm g}\,{\rm cm}^{-2},
\end{equation}
where $u = \ln E$ [GeV] and the coefficients $P_i$ are listed in Table 1.
The hadronization algorithm that translates the parton strings produced 
during the scattering process into ordinary particles, remains the same.

In the simulation, several sets of protons with $E = 3 \times 10^{20}$ eV 
were injected at 100 km above sea level (a.s.l.).
The sample was uniformly spread in the interval of 
0$^{\circ}$ to 50$^{\circ}$ zenith angle at
the top of the atmosphere. All shower particles with energies above the 
following thresholds were
tracked: 750 keV for gammas, 900 keV for electrons and positrons, 10
MeV for muons, 60 MeV for mesons and 120 MeV for nucleons.
The results of these simulations were processed with the help of the 
{\sc aires} analysis package.

The atmospheric depth $X_{\rm max}$ at which
the shower reaches its maximum number of secondary particles is the
standard observable to describe the speed of the shower development.
The charged multiplicity, essentially electrons and positrons, is used 
to determine the number of charged particles and the location of the shower 
maximum by means of 4-parameter fits to the Gaisser-Hillas function 
 \cite{gaisser-hillas},
\begin{equation}
N^{\rm ch} (X) = N^{\rm ch}_{\rm max} \left(
\frac{X - X_0}{X_{\rm max}-X_0}\right)^{[(X_{\rm max} - X_0)/\lambda]}\,\, 
\exp\left\{\frac{X_{\rm max} - X}{\lambda}\right\},
\,\,\, X \geq X_0,
\end{equation}
where $X_{\rm max}$, 
$N^{\rm ch}_{\rm max}$, $\lambda$, and $X_0$ are the free parameters 
to be adjusted \cite{steve}. 
Shown in Fig. 2 are the resultant $X_{\rm max}$ distributions of proton 
showers with $3 \times 10^{20}$ eV and primary zenith 
angle 43.9$^\circ$.\footnote{This is the primary zenith angle of the 
Fly's Eye event \cite{FE}.}
The tails ($X_{\rm max} > 900$ g/cm$^{2}$) of these distributions
were fitted 
with exponentials ($\alpha\,e^{-\beta\,X_{\rm max}}$), floating
both the normalisation $\alpha$ and the exponent in the fit.
The resulting 
parameters are: $\beta = 2.6 \pm 0.1 \times 10^{-2}$ cm$^2$/g 
for the 4-dimensional  
case, and $\beta = 2.9 \pm 0.1 \times 10^{-2}$ cm$^2$/g for 
the $(4+n)$-dimensional case. A statistically significant 
difference between the two approaches arises in the tail of the 
distribution. This is because the depth of  
such penetrating showers increasingly reflects that of the 
first interaction \cite{gaisser}.
Results of the fits to the $X_{\rm max}$ distributions generated by 
applying progressively less restricted data cuts (distances near the peak)
lead to exponential slopes that within the error are consistent with one 
another.

\begin{figure}
\label{hamburg1}
\begin{center}
\epsfig{file=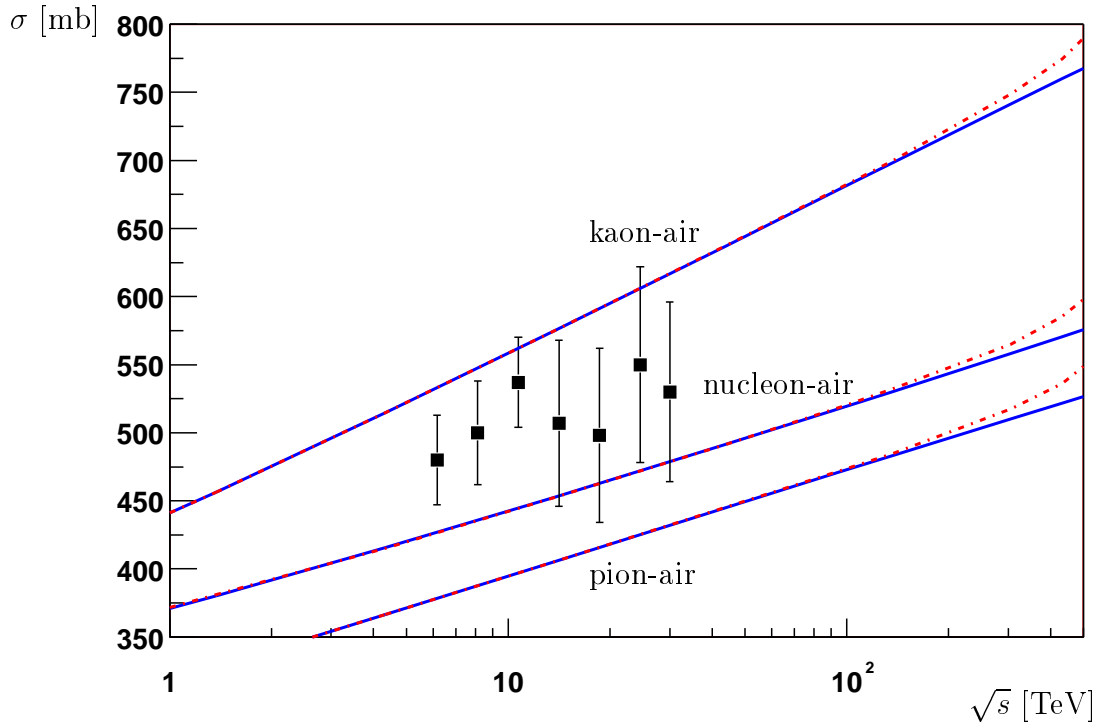,width=15.cm,clip=} 
\caption{Inelastic cross sections as a function of the c.m. energy.
The solid line stands for the usual 4-dimensional 
cross section of {\sc qgsjet}, whereas the 
dashed line represents corrections coming from the virtual graviton 
exchange.
We also show in the figure experimental points of the inelastic $p$-air 
cross section as 
observed by different cosmic ray experiments \cite{cs}.}
\end{center}
\end{figure}
\begin{table}
\caption{Coefficients for mean free path parametrization, $M_* = 1$ TeV}
\begin{center}
\begin{tabular}{ccccccccc}
\hline\hline 
particle & $P_1$ & $P_2$ & $P_3$ & $P_4$ & $P_5$ & $P_6$ & $P_7$ 
& $P_8$
\\ \hline 
nucleons & -59.852 & -1916.4 & -25.508 & 3.2875 & 925.76 & 69.860 
& -0.089103 & 
-0.12169 \\
pions & -70.680 & -1500.1 & -26.015 & 2.7753 & 585.44 & 69.425 & 
0.36761 & -0.12197 \\
kaons & -84.984 & -953.40 & -23.677 & 2.0865 & 262.26 & 54.498 
& 0.70365 & -0.10659\\
\hline \hline
\end{tabular}
\end{center}
\end{table}

\begin{figure}
\label{hamburg4}
\begin{center}
\epsfig{file=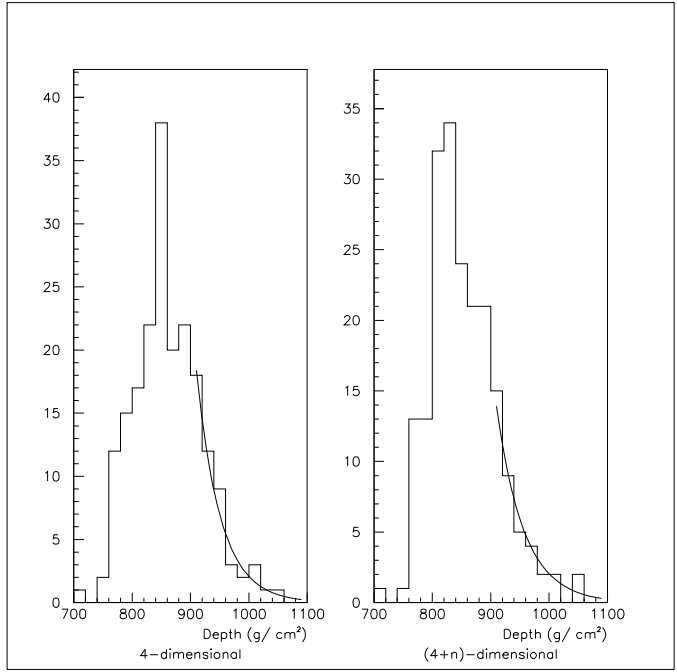,width=9.cm,clip=} 
\caption{Distributions of $X_{\rm max}$.}
\end{center}
\end{figure}
\begin{figure}
\label{hamburg2}
\begin{center}
\epsfig{file=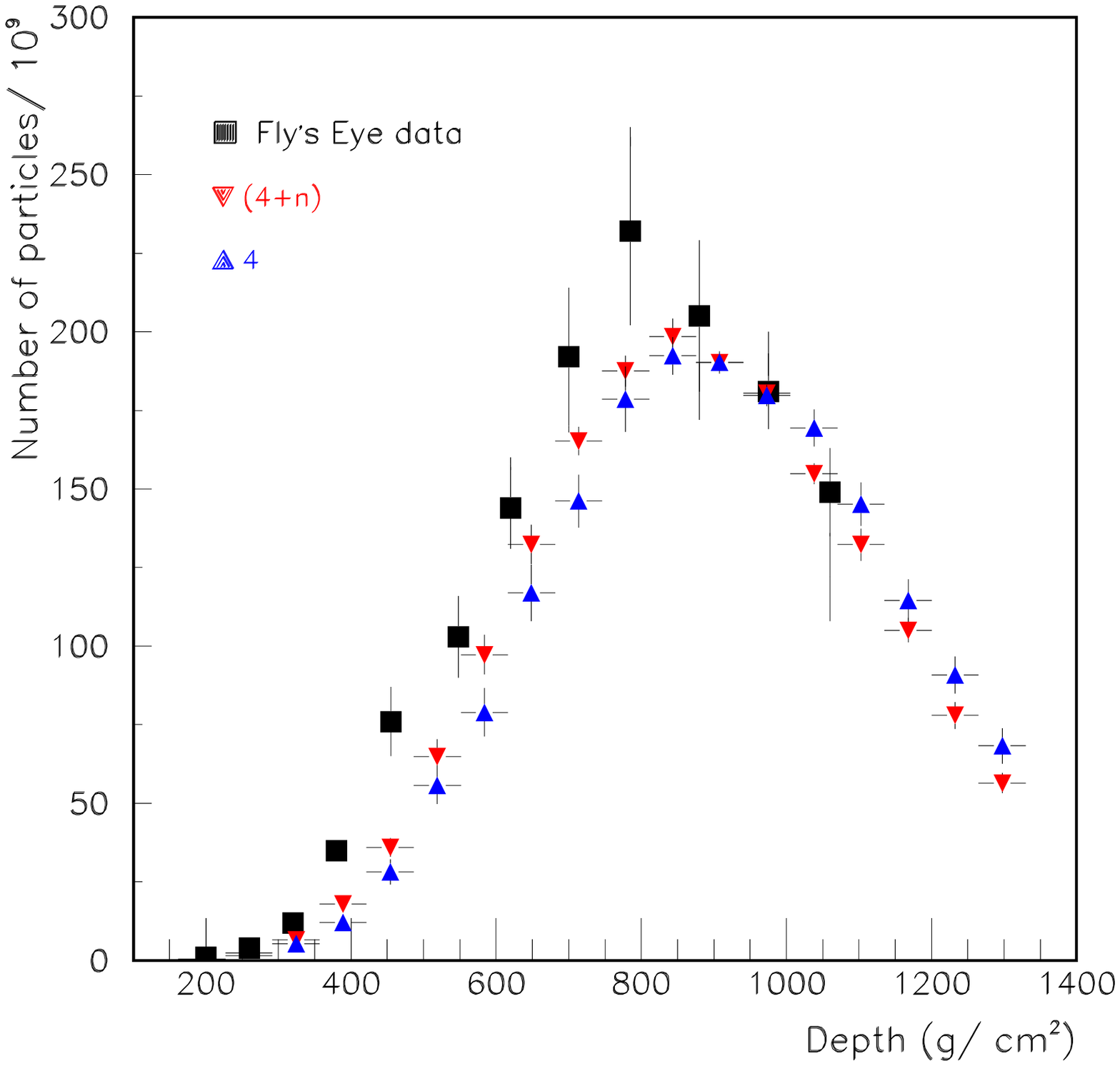,width=13.cm,clip=} 
\caption{Atmospheric cascade development of proton showers 
($E= 3 \times 10^{20}$ eV),  superimposed over the Fly's Eye data.
The error bars in the simulated curves indicate RMS 
fluctuations of the means.}
\end{center}
\end{figure}

In Fig. 3 we show the longitudinal developments of proton showers 
superimposed over the experimental data of the world's highest energy 
cosmic ray shower observed to date \cite{FE}. We selected from our shower sample those
with a 
primary zenith angle of $43.9^\circ$, setting the observation level at 
850 m a.s.l and with geomagnetic field specific for the Fly's Eye site.
Although at the same total energy a shower that takes into account 
the virtual graviton exchange develops faster than that modelled with 
unmodified {\sc qgsjet}, as expected from our previous analysis, the 
differences in the position of $X_{\rm max}$ fall within the errors. 
However, there are visible deviations in the evolution of the charged 
multiplicity. To estimate the amount of departure from the standard 
4-dimensional scenario we analyzed the data by means of a $\chi^2$ 
test \cite{pdg}. We assume that the set of measured values by Fly's Eye are 
uncorrelated (any depth measurement is independent of any other), and 
make use of the quantity 
\begin{equation}
\chi^2 \equiv \sum_{j=1}^q \frac{|x_j - \alpha_j|^2}{\sigma_{x_j}^2},
\end{equation}
where $q$ is the total number of points in the analysis, 
$\sigma_{x_j}$ is the error on the $x_j$th coordinate, $x_j$ is the measured 
value of the coordinate, and $\alpha_j$ the (hypothetical) true value of the 
coordinate. The obtained results are $\chi^2_{4}/{\rm DOF} = 324.89/12$, 
$\chi^2_{(4+n)}/{\rm DOF} = 200.52/12$. If in the future the situation 
should arise that one can be confident that the hadronic interactions 
are correctly modeled, then it will be necessary to
carry out a more sophisticated statistical analysis which, for example,
accounts for the non-Gaussian distributions.
                
It is also interesting to inspect whether KK graviton exchange has any 
influence on the particle densities 
at ground level. A summary of the ground lateral distributions of 
proton showers at vertical incidence ($E = 3 \times 10^{20}$ eV) 
is reported in Table 2. Here, the 
ground array was located at 875 g/cm$^2$ and the
magnetic field was set to reproduce that prevailing upon the Auger experiment.
The ratio between the mean density of charged particles 
$\rho_{\rm ch}^{4}/\rho_{\rm ch}^{(4+n)}$ is a monotonically 
decreasing function of the 
distance to the shower core $R$. Nevertheless, one should note that 
within the error limits the ratio, $\rho_j^{4}/\rho_j^{(4+n)}$ 
($j =\mu^{\pm},\,e^\pm,\,\gamma$, all charged particles), is always 
consistent with 1. Thus, we deduce that the possible signatures 
of KK emission are entirely hidden when the shower front reaches the ground.   
Furthermore, the competition of decay and interaction of the 
first generations of mesons propagating in a medium with varying density 
profile,
is of particular relevance in the indirect analysis of data collected by 
ground arrays. As can be seen in Fig. 4, the ratio 
$\rho_j^{4}/\rho_j^{(4+n)}$ would generally depend on the primary 
zenith angle.

\begin{table}
\caption{Particle densities [m$^{-2}$]. The errors indicate the RMS fluctuations.}
\begin{center}
\begin{tabular}{ccccc}
\hline\hline 
$\,\,$ $4$ $\,\,$&   & $R$ = 50 m & $R$ = 500 m & $R$ = 1000 m \\        
$\,\,$ $\rho_{\rm ch}$ $\,\,$ & & $269.05 \pm 19.9 \times 10^{4}$ & $14.41 
\pm 1.51 \times 10^{2}$ & $ 82.38 \pm 26.10 \times 10^0$ \\
$ \,\,$ $\rho_{\mu^{\pm}}$  $\,\,$ & & 
$138.02 \pm 5.37 \times 10^2$ &  $19.43 \pm 0.86 \times 10^1$ 
& $19.71 \pm 01.32 \times 10^0 $\\
$\,\,$ $\rho_{e^{\pm}}$ $\,\,$ & &  $267.56 \pm 19.9 \times 10^{4}$ & 
$12.41 \pm 1.50 \times10^2$ &  $62.46 \pm 26.00 \times 10^0$ \\
$\,\,$ $\rho_\gamma$ $\,\,$ & & $148.85 \pm 3.90 \times 10^{5}$ & 
$25.02 \pm 1.04 \times 10^{3}$ & $91.73 \pm 10.60 \times 10^{1}$ \\
\hline
$\,\,$$(4+n)$ $\,\,$&  & $R$ = 50 m & $R$ = 500 m & $R$ = 1000 m \\
$\,\,$ $\rho_{\rm ch}$ $\,\,$ &  &$ 245.98 \pm 7.97 \times 10^{4}$   &   
$ 15.09 \pm 1.58 \times 10^2$ & $10.99 \pm 02.08 \times 10^{1}$ \\
$ \,\,$ $\rho_{\mu^{\pm}}$ $\,\,$ & & $ 136.85 \pm 6.26 \times 10^{2}$ 
& $19.43 \pm 0.90 \times 10^1$ & $ 21.61 \pm 01.63 \times 10^0$ \\
$\,\,$ $\rho_{e^{\pm}}$ $\,\,$ & & $244.49 \pm 7.93 \times 10^{4}$  
& $13.11 \pm 1.56 \times 10^2$ &$ 88.28 \pm 21.10 \times 10^0$ \\ 
$\,\,$ $\rho_\gamma$ $\,\,$& & $150.96 \pm 4.46 \times 10^{5}$ &
$27.10 \pm 1.94 \times 10^{3}$ & $10.40 \pm 01.42 \times 10^2$\\
\hline\hline
\end{tabular}
\end{center}
\end{table}
\begin{figure}
\label{hamburg3}
\begin{center}
\epsfig{file=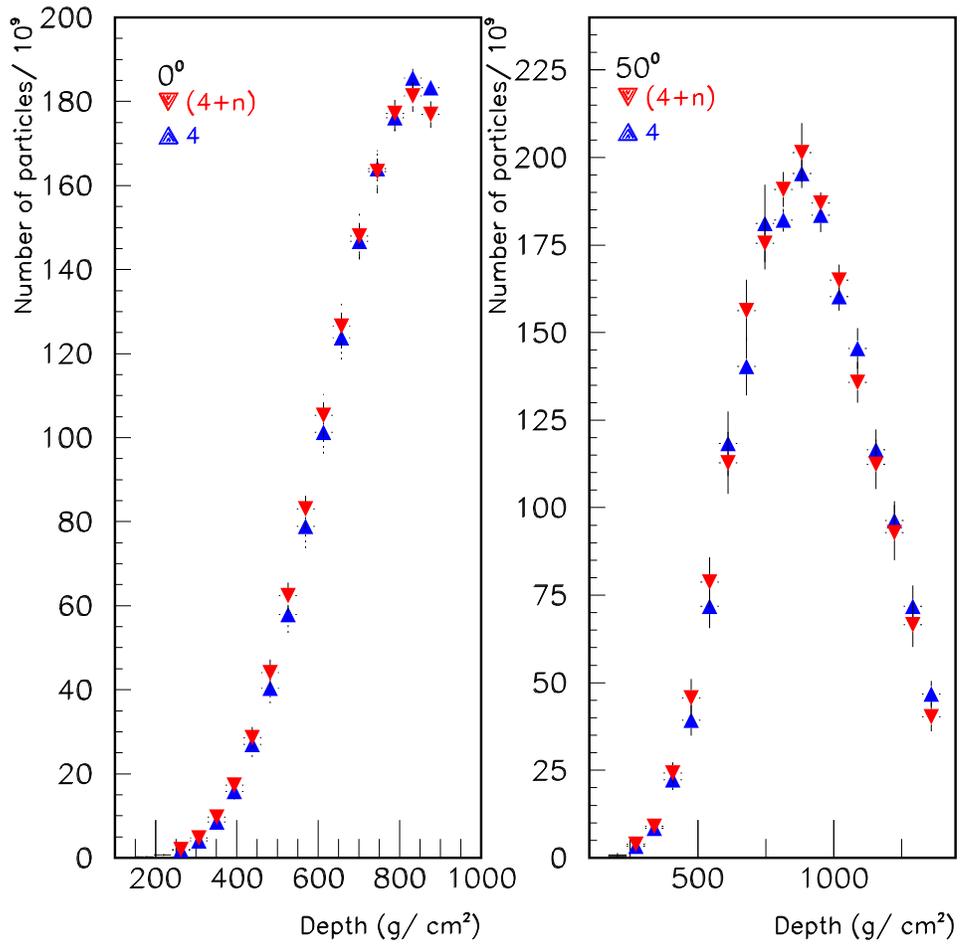,width=13.cm,clip=} 
\caption{Atmospheric cascade developments of proton showers 
 for extreme primary zenith angles ($0^\circ$ and $50^\circ$) and $E= 3 \times 10^{20}$ eV.
The error bars indicate RMS 
fluctuations of the means.}
\end{center}
\end{figure}

Putting all this together, {\it KK-graviton exchange offers a viable 
mechanism to reduce by around 6\% the mean free path of ultra high energy
($E> 5 \times 10^{19}$ eV) hadrons in the atmosphere}.

\section{Conclusion}

\hfill

Theories with large compact dimensions and TeV-scale quantum gravity 
represent a radical departure from previous fundamental particle physics.
If these scenarii have some truth, the scattering phenomenology 
above collider energies would be quite distinct from SM expectations.
In particular, the  exchange of KK towers of gravitons
leads to a modification of  SM hadronic cross sections at $s^{1/2} > 100$ TeV.
Extremely high energy cosmic rays that 
impinge on stationary nucleons at the top of the atmosphere start 
chain reactions where the c.m. energy can be as high as 500 TeV.
It is therefore instructive to explore KK exchange sensitivity within 
the entire average profile of the air shower. In this paper we have 
contributed a few results to this question.
We have shown that the exchange of KK gravitons could affect the rate 
of development of atmospheric cascades initiated by protons. 
For primary energies above $3 \times 10^{20}$ eV, the effects are 
statistically significant and can thus be observed by fluorescence 
detectors \cite{l}. We have also proved that the 
footprints left by the $(4+n)$-dimensional gravitons become washed out 
as the shower front gets closer to the ground and, in general, 
cannot be traced back with surface array data. The details of our 
analysis should be 
treated with some caution since 
they may be sensitive to the hadronic interaction model used. The overall 
conclusion, however, should remain the same.

\section*{Acknowledgments}

This work was partially supported by CONICET (Argentina) and the 
National Science Foundation.


\begin{thebibliography}{99}
\bibitem{g} Assuming that the higher dimensional theory at short distance is a string theory, one expects that the fundamental string scale $M_s$ and $M_*$ are not too different. A perturbative expectation is that $M_s \sim g_s M_*$, where $g_s^2$ is the gauge coupling at the string scale, of order 
$g^2_s/4\pi \sim 0.04$.
\bibitem{ouo} For $\ell/r_c > M_s$ we would expect higher order 
quantum gravity and stringy effects to become important providing 
some sort of a natural cut--off in the field theory.
\bibitem{csec} Since we expect interference terms, Eq. (\ref{cs}) 
must be understood as a lower bound on the effect. Note that a lower bound on 
the effect does not necessarily imply a lower bound on the cross section.
\bibitem{M_air} The mean molecular weight corresponding to an atomic 
mixture of 78\% N, 21.05\% O, 0.47 \% Ar and 0.03 \% other elements
is $M_{\rm air} = 28.966$. Then $m_{\rm air} \approx 2.43 
\times 10^{-23}$ g.
\bibitem{steve} The parameter $\lambda$ is externally fixed at 70 g/cm$^2$.
This is the usual procedure in {\sc aires}.
\end{thebibliography}
\end{document}